\documentstyle[aps,preprint,epsfig]{revtex}
\newcommand{\imag}{{\rm i}}

\tighten
\begin{document}
\draft
\author{J. P. Dewitz and W. H\"ubner}
\address{Max Planck Institut f\"ur Mikrostrukturphysik, Weinberg
  2, D-06120 Halle/Saale, Germany}
\date{\today}
\title{Nonlinear Magneto-Optics of freestanding Fe monolayers from first principles}
\maketitle

\begin{abstract}
The nonlinear magneto-optical Kerr-effect (NOLIMOKE) is determined for
freestanding Fe monolayers with several in-plane structures from
first principles. Based on the theory of nonlinear
magneto-optics by H\"ubner and Bennemann~[Phys. Rev. B, {\bf 40}, 5973 (1989)] we calculate the nonlinear
susceptibilities of the monolayers using the {\em ab initio}
FLAPW-method WIEN95 with the additional
implementation of spin-orbit coupling and the
calculation of the dipole transition matrix elements appropriate for
freestanding monolayers. We present results for the spectral dependence of the nonlinear susceptibility tensor elements and the resulting intensities and Kerr angles. Special emphasize is put on the effects of structural changes such as the variation of the lattice constant and different surface orientations. The influence of spin-orbit coupling on the tensor elements for different magnetization directions is presented as well as the azimuthal dependence of the intensities generated by several low index surfaces, showing the pronounced sensitivity of second harmonic generation to lateral structural changes as well as magnetic properties even in the monolayer range.
\end{abstract}
\section{Introduction}
The nonlinear magneto-optical Kerr-effect (NOLIMOKE) combines the surface and interface sensitivity of second harmonic generation (SHG) with the sensitivity on magnetic ordering, which is much stronger here than in linear optics~\cite{Pusto:94.prb,Rasing:96.jmmm}.
The existence and the detectability of the nonlinear magneto-optical effect was shown independently in 1989 by Pan, Wei and Shen~\cite{PWS} and H\"ubner and Bennemann~\cite{Huebner:89}. Whereas the former work was based on a group theoretical classification of the nonlinear susceptibility tensor $\chi_{ijk}^{(2)}$, the latter used an electronic structure calculation to predict the nonlinear Kerr-effect. During the last years significant improvements have been made. The group theoretical analysis was extended to antiferromagnetic surfaces~\cite{Daehn:96}. Also several extensions of the electronic structure calculations have been performed, such as the calculation of the nonlinear magneto-optical response of Ni within a tight binding scheme~\cite{Huebner:90} and of Fe layers and films using a FP-LMTO method~\cite{Pusto:ZPB}. In these papers spin-orbit coupling (SOC) was treated perturbatively and only recently~\cite{Dewitz:soc} SOC was included from first principles to obtain the nonlinear susceptibility of Fe monolayers for several in-plane structures. Nevertheless, no complete theory was available combining both the group theoretical analysis and first principles electronic structure calculations, since up to now the dipole transition matrix elements have not been determined from first principles.

In this paper we present a calculation of the nonlinear magneto optics for freestanding Fe monolayers, which obtains all fundamental aspects of magneto-optics, the electronic structure including spin-polarization, SOC, and the dipole transition matrix elements from first principles. The symmetry aspects are completely determined by the symmetries of the wavefunctions.
\section{Theory}
Our calculations are based on the formula for the nonlinear susceptibility tensor
\begin{eqnarray}
  \label{eq:suszep1}
\!\!\!\!\!\!\!\!\!\!\!\!\!\!\!\!\!\!\!\!\!\!\!\!\chi^{(2)}_{ijk}\left(2{\bf q},2
\omega\right) = e^3\sum_{{\bf k},l,l',l''}
\left\{\left<{\bf k}+2{\bf q},l''|i|{\bf k},l\right>\left<{\bf
    k},l|j|{\bf k+q},l'\right>\left<{\bf k+q},l'|k|{\bf k}+2{\bf
    q},l''\right>\right.\quad\\
\!\!\!\!\!\!\!\!\!\!\!\!\!\!\!\!\!\!\!\!\!\!\!\!\!\!\!\!\!\!\!\!
\left.\times\frac{\frac{f(E_{{\bf
          k}+2{\bf q},l''})-f(E_{{\bf k}+{\bf
          q},l'})}{E_{{\bf k}+2{\bf
          q},l''}-E_{{\bf k}+{\bf
          q},l'}-\hbar\omega+i\hbar\alpha_1}-
    \frac{f(E_{{\bf k}+{\bf q},l'})- f(E_{{\bf
          k},l})}{E_{{\bf k}+{\bf q},l'}-
      E_{{\bf k},l}-\hbar\omega+i\hbar\alpha_1}}{E_{{\bf k}+
      2{\bf q},l''}-E_{{\bf k}l}-
    2\hbar\omega+i2\hbar\alpha_1}\right.\qquad\nonumber \\
\!\!\!\!\!\!\!\!\!\!\!\!\!\!\!\!\!\!
\left.\times\left[1+4\pi{\rm e}^2\sum_{ij}m_im_j\sum_{{\bf
      k},l,l''}\left<{\bf k}l\left|{\bf r}_i\right|{\bf k}+2{\bf
        q},l''\right>\left<{\bf k}+2{\bf q},l''\left|{\bf
    r}_j\right|{\bf k},l\right>\frac{f(E_{{\bf k}+2{\bf
        q},l''})-f(E_{{\bf k},l})}{E_{{\bf k}+2{\bf q},l''}-E_{{\bf
        k}l}-2\hbar\omega+2{\rm i}\hbar\alpha}\right]^{-1}\right\}\;.\nonumber
\end{eqnarray}
derived in~\cite{Huebner:89}. Therein {\bf q} is the momentum of the incident light, $\alpha$ is a damping constant simulating a finite life time and chosen as 0.2 eV, $f(E_{{\bf k},l})$ denotes the Fermi-distribution and $m_i$ is the $i$-th direction cosine of {\bf q} with respect to the plane of incidence. The energy bands $E_{{\bf k},l}$ and the wavefunctions $\left|{\bf k},l\right>$ are determined using the {\em ab initio} FP-LAPW method WIEN95~\cite{wien95}. SOC is included as described in~\cite{Dewitz:soc}. Within this treatment the wavefunction reveal the correct symmetry of the system including SOC. Thus, by determining the dipole transition matrix elements $\left<{\bf k},l|i|{\bf k}',l'\right>$ from the wavefunctions, we obtain the correct form of the susceptibility tensor with respect to symmetry as derived in~\cite{PWS} from first principles. The numerical differences between group theoretically allowed and forbidden tensor elements is about 12 orders of magnitude. The precise treatment will be described elsewhere~\cite{jdfinal}. 

Since, at this stage of the investigations, we are mainly interested in the effects of lateral structural changes on the magneto-optical spectra, we calculate only freestanding monolayers neglecting the substrate. The absence of a substrate enables us to choose arbitrary in-plane structures and reduces the required computer time to a reasonable value. Of course the substrate cannot be neglected for a quantitative determination of the magneto-optical response and will be included in further studies.

Within the electric-dipole approximation the presence of inversion symmetry breaking is necessary for the occurrence of SHG. Thus, in the case of freestanding monolayers, we have to introduce an artificial symmetry breaking perpendicular to the surface.
 This is done by integrating the dipole transition matrix elements in the upper half of the unit cell only, which effectively sets the wavefunctions in the lower half volume equal to a constant. Roughly speaking, this simulates a substrate with a constant charge density. In doing so we also solve the general problem of treating semi infinite systems like film or layer structures with a {\bf k}-space method like WIEN95, which uses threedimensional translational invariance.

The SHG intensities are then calculated from the nonlinear susceptibilities using the expression for the second harmonic generated field given in~\cite{Huebner:Boehmer}.
\section{Results}
\begin{figure}
\caption{Spectral dependence of the imaginary part of the nonlinear susceptibility tensor elements of a Fe(001) monolayer for different lattice constants $a$ given in \AA. The magnetization is out-of-plane ({\bf M} $\parallel$ {\bf z}). The lattice constant is varied between 2.4 and 2.76 \AA.}
\label{fig:tensor}
\caption{Spectral dependence of the linear and nonlinear Kerr angle of a Fe(001) monolayer with out-of-plane magnetization. The lattice constant is varied between 2.4 and 2.76 \AA.}
\label{fig:kerr}
\caption{Spectral dependence of the nonmagnetic tensor elements of a Fe(001) monolayer with out-of-plane and in-plane magnetization. The weak dependence reveals the SOC dependence of second order.}
\label{fig:soc}
\caption{Azimuthal dependence of the nonlinear intensities with $p$ polarization of the incident and generated light for the Fe(001), Fe(110) and Fe(111) monolayer with out-of-plane and in-plane magnetization.}
\label{fig:azi}
\end{figure}
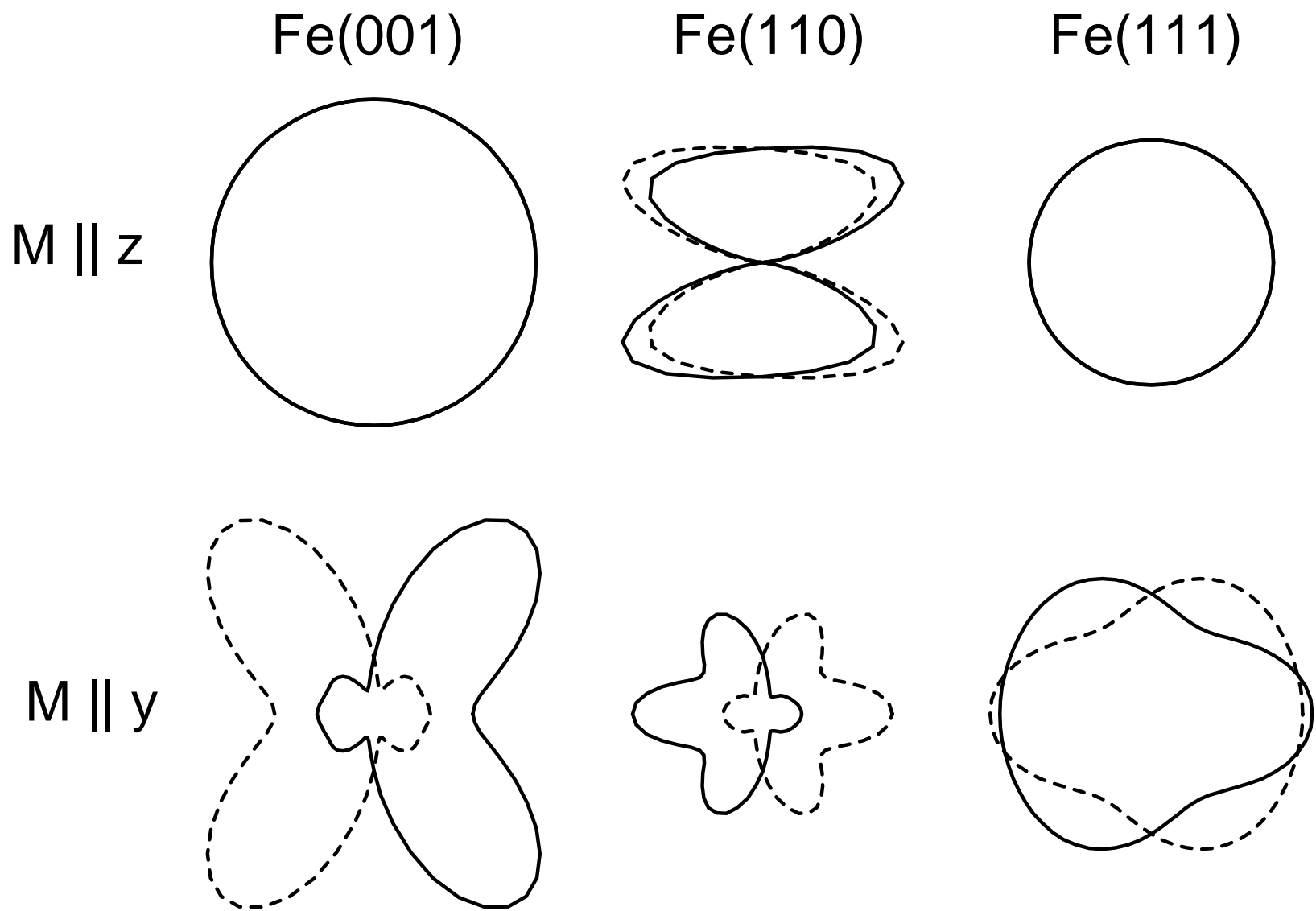
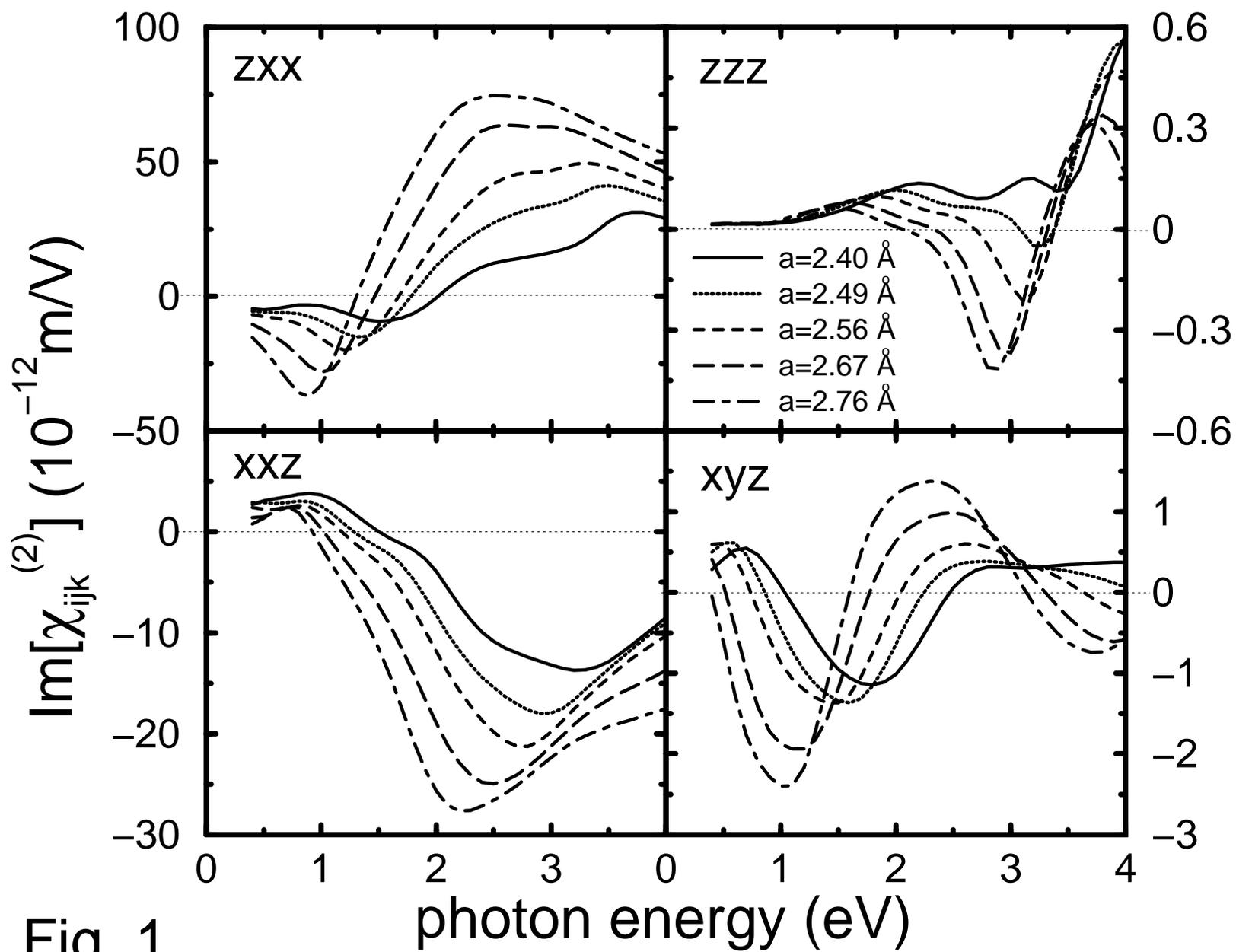
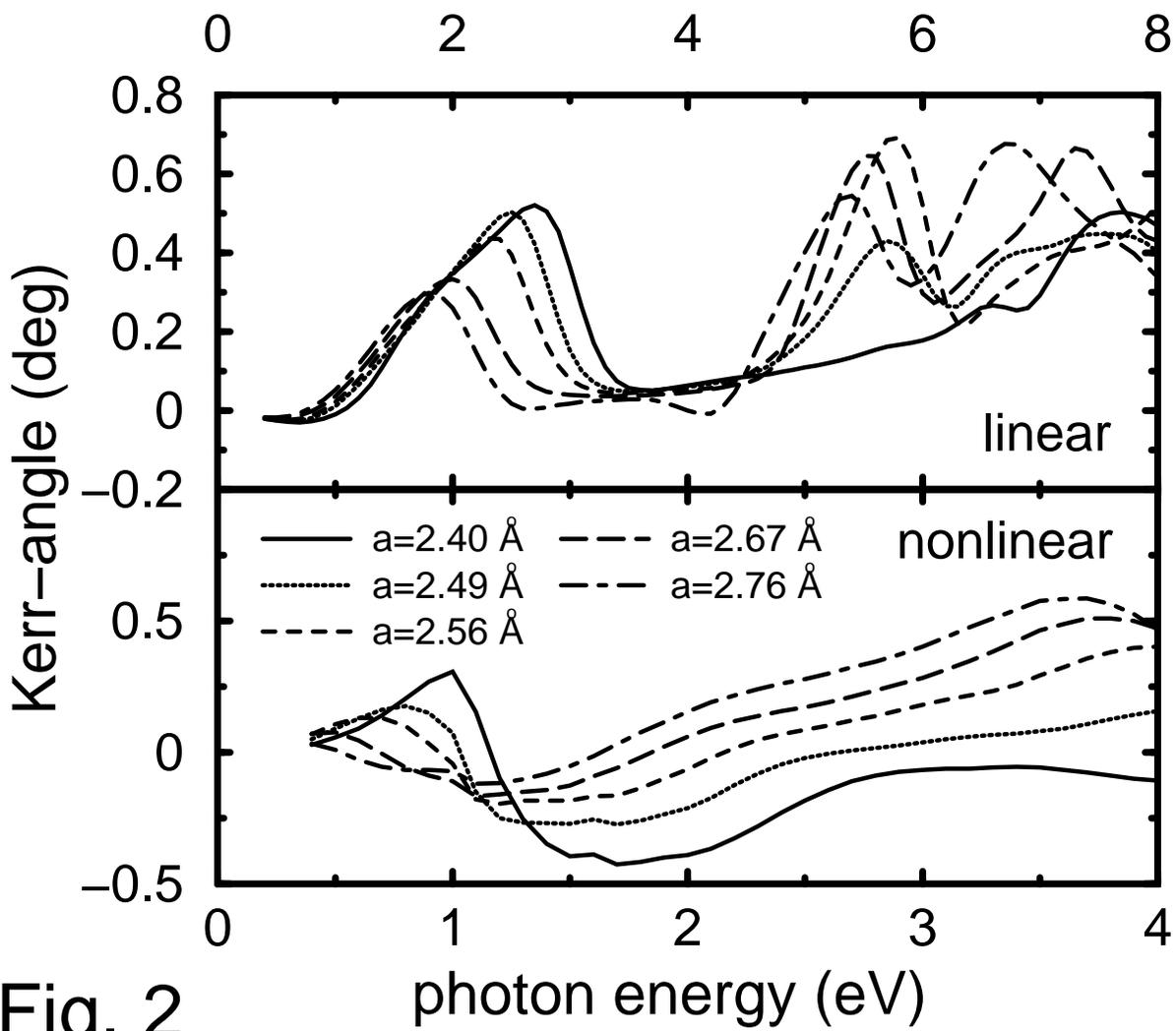
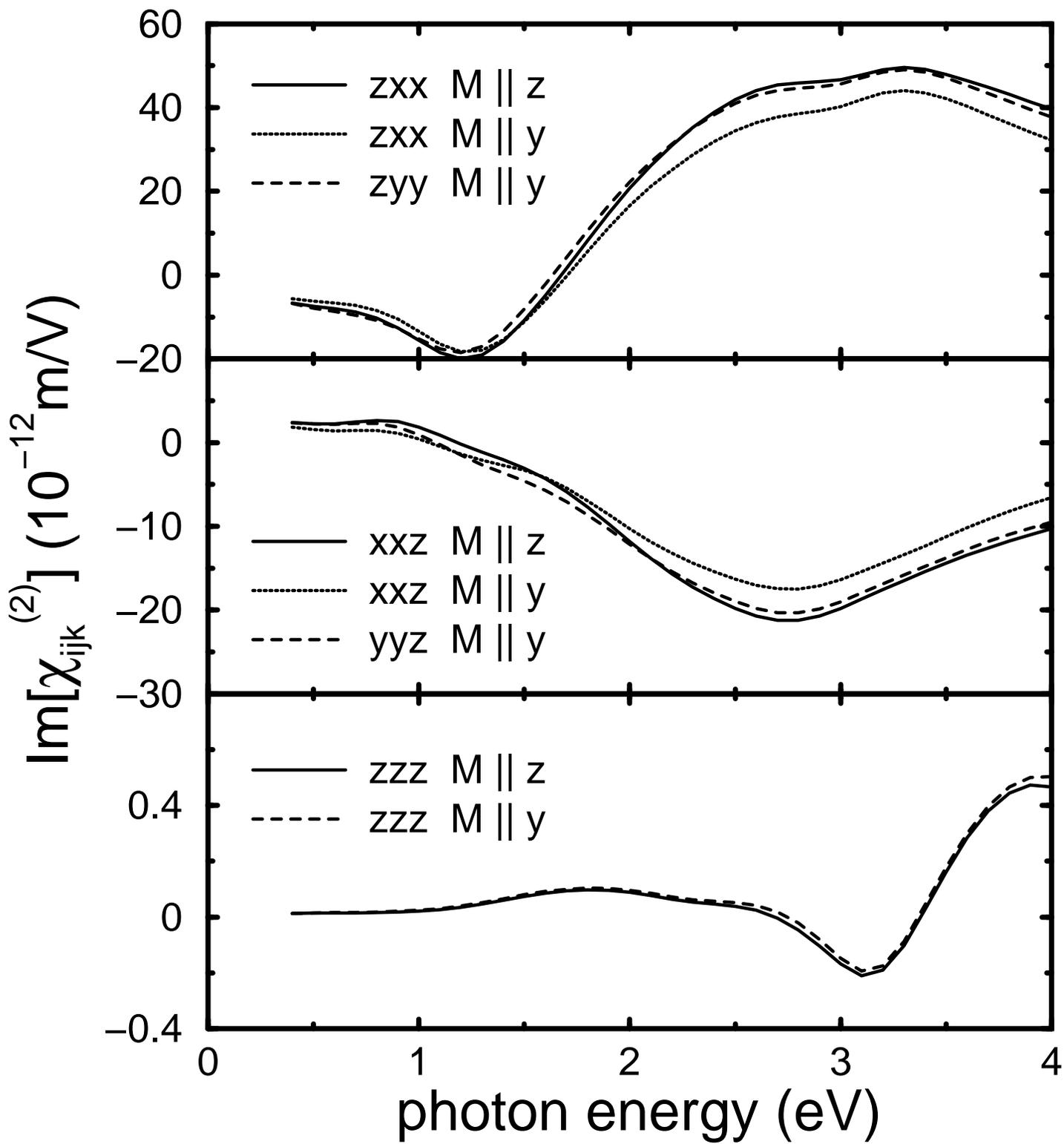

Fig.~\ref{fig:tensor} shows the spectral dependence of the imaginary part of the four independent tensor elements of the nonlinear susceptibility $\chi^{(2)}_{ijk}$ obtained from a Fe(001) monolayer with out of plane magnetization ({\bf M} $\parallel$ {\bf z}) derived from a fcc lattice. The in-plane lattice constant is varied from 2.4 \AA~to 2.76 \AA~(Cu has $a$=2.56 \AA), simulating different substrates . For this system the nonlinear susceptibility tensor has the following form:
\begin{equation}\chi^{(2)}=
\left(\begin{array}{ccc|ccc}
    0  &  0  &  0  & xyz^- & xxz^+ &  0  \\
    0  &  0  &  0  & xxz^+ & yxz^- &  0  \\
   zxx^+ & zxx^+ & zzz^+ &  0  &  0  &  0  \end{array}\right)\;.
\end{equation}
There are four independent tensor element $zxx^+,zzz^+,xxz^+$ and $xyz^-$. The first three do not change their sign under magnetization reversal ($^+$ superscript) and thus will be called ``nonmagnetic''~\cite{bem1}, whereas the $xyz$-element changes its sign ($^-$ superscript) and thus are termed ``magnetic''. The tensor elements show a clear dependence on the lattice constant, as was already found in calculations with constant matrix elements~\cite{Pusto:ZPB,Dewitz:soc}. Zeros in the spectra are shifted to larger energies for smaller lattice constants, maxima have a larger magnitude for larger lattice constants. Of course, the spectral dependence becomes much more complicated with the inclusion of the matrix elements, since the latter affect both the width and absolute values of the resonances. The nonlinear Kerr angle $\Phi_{\rm K,p}^{(2)}$ in the polar configuration with $p$ input polarization is given by
\begin{equation}
\label{eq:pkerrwnoli}
\Phi_{\rm K,p}^{(2)}+\imag \varepsilon_{\rm K,p}^{(2)} = \frac{A_s f_c f_s \chi^{(2)}_{xyz}}{A_p [N^2F_s(f_c^2\chi_{zxx}^{(2)} + f_s^2\chi_{zzz}^{(2)})+2F_cf_cf_s\chi^{(2)}_{xxz}]}
\end{equation} 
where $f_s,c$ are Fresnel coefficients of the incident field, $A_{p,s}$ and $F_{c,s}$ are transmission and Fresnel coefficients of the generated field and $N$ is the refractive index of the substrate material. $\varepsilon_{\rm K}^{(p)}$ is the ellipticity and an approximation for small angles is used. From Fig.~\ref{fig:tensor} it is clear that the contribution of the $\chi^{(2)}_{zzz}$ and $\chi^{(2)}_{xxz}$ elements can be neglected because of their magnitude. Thus the nonlinear Kerr angle is directly proportional to the ratio of the $\chi^{(2)}_{xyz}$ and the $\chi^{(2)}_{zxx}$ elements weighted by the transmission and Fresnel coefficients (in the numerical calculation, however all tensorelements are included). Though the spectral dependence of latter is rather weak in this energy range compared to the tensor elements, a clear relation between the spectral dependence of the Kerr angle and the tensor elements is not obvious. Nevertheless both the spectral dependence of the Kerr angles and the relevant tensor elements for the different lattice constants (see Fig.~\ref{fig:tensor}) show some common features. Maxima around 1~eV are followed by a crossing of the curves between 1 and 1.5~eV. The spectral dependence of the linear Kerr angles in the upper part of Fig.~\ref{fig:kerr} shows a similar feature at twice the energy but exhibits much more structure for higher photon energies. More detailed investigations of the involved complex quantities show that the spectral dependence of the nonlinear Kerr angles is mainly ruled by the magnetic tensor elements. Fig.~\ref{fig:kerr} shows that in our calculations the linear and nonlinear Kerr angles are of the same order of magnitude, which does not agree with the experimental findings~\cite{VollmerBuch}. This is due to the specific form of the artificial inversion symmetry breaking performed here, which also underestimates the size of the $zzz$ tensor element. Note, the present study focuses on the structural effects on the nonlinear Kerr spectra. This problem only occurs in the case of freestanding monolayers and can be overcome by including substrate effects and using unit cells for the electronic structure calculations in which inversion symmetry is explicitly broken~\cite{jdfinal}. Furthermore the size of the tensor elements in Fig.~\ref{fig:tensor} is too large compared to previous estimates~\cite{Pusto:93}. This can be a problem of the neglected screening effects at the surface. 

Fig.~\ref{fig:soc} shows the differences in the spectral dependence of the nonmagnetic tensor elements of the Fe(001) monolayer with a lattice constant $a$=2.56 \AA~for in-plane ({\bf M} $\parallel$ {\bf y}) and out of plane ({\bf M} $\parallel$ {\bf z}) magnetization. Since the two dimensional unit cell of the (001) surface is a square lattice, the $x$ and $y$ direction are equivalent for out-of-plane magnetization, and thus $\chi^{(2)}_{zxx}=\chi^{(2)}_{zyy}$. For in-plane magnetization the $x$ and $y$ directions are no longer equivalent and thus $\chi^{(2)}_{zxx}\neq\chi^{(2)}_{zyy}$. The effect of this symmetry breaking on the nonmagnetic tensor elements is shown in Fig.~\ref{fig:soc}, yielding only a slight dependence on the magnetization direction. This is expected from the fact that nonmagnetic tensor elements are only of second order in SOC~\cite{Kittel}.

One advantage of SHG compared to linear optics is the enhanced symmetry resolution due to the third rank tensor $\chi_{ijk}^{(2)}$ involved in SHG compared to the second rank tensor $\chi_{ij}^{(1)}$ in linear optics. The in-plane symmetry resolution can be monitored by calculating the azimuthal dependence of the intensities. This is displayed in Fig.~\ref{fig:azi}, where the azimuthal dependence of the intensities with $p_{\rm in}$ and $p_{\rm out}$ polarization is plotted for the Fe(001), Fe(110), and Fe(111) monolayer with in-plane and out-of-plane magnetization. SHG can resolve up to three-fold rotational symmetries. Thus, for {\bf M} $\parallel$ {\bf z}, the SHG signal of the square lattice Fe(001) which has four-fold symmetry in this configuration show no azimuthal dependence. Also the Fe(111) surface with {\bf M} $\parallel$ {\bf z} shows no azimuthal dependence since, in the absence of a substrate, this monolayer has a six-fold symmetry. The curve for the Fe(110) surface reveals its twofold symmetry. The dashed line indicates the intensities for inverted magnetization direction, showing the effect of broken symmetries due to the magnetism and SOC. For an in-plane magnetization this effect is increased. The magnetization shifts the SHG intensities in the direction perpendicular to the magnetization. Also the well-known effect of a large magnetic contrast in the intensities in the transversal configuration is reproduced by our {\em ab initio} theory. Finally, it is clearly visible that the in-plane magnetization destroys the circular dependence of the intensities in the case of the Fe(001) and Fe(111) monolayers.
\section{Summary and Discussion}
In summary we showed results on the nonlinear magneto optical effect of Fe monolayers as obtained from first principles calculations. The dipole transition matrix elements were computed from the wavefunctions. The inversion symmetry has been artificially broken to obtain SHG from the freestanding monolayers. The symmetries of the system including SOC are completely reflected by the wavefunctions yielding the correct form of the nonlinear susceptibility tensor known from group theoretical classifications~\cite{PWS}. The tensor elements as well as the resulting Kerr angle show a clear dependence on the lattice constant in agreement with previous results using constant matrix elements. The spectral dependence of the nonlinear Kerr angle shows up as a complex superposition of the involved tensor elements and transmission and Fresnel coefficients. The precise inclusion of SOC reveals a weak dependence of the nonmagnetic tensor elements on the magnetization direction. The high sensitivity of SHG to the in-plane structure was shown for the Fe(001), Fe(110), and Fe(111) monolayers, showing also the important influence of the magnetic tensor elements.

We acknowledge financial support by TMR Network NOMOKE contract No. FMRX-CT96-0015.

\end{document}